\documentclass{jfm}
\usepackage[english]{babel}

\usepackage{graphicx}
\usepackage{natbib}
\usepackage{color}

\def\upi{\pi}

\def\Uv{{\boldsymbol U}}  
\def\xv{{\boldsymbol x}}  
\def\xiv{{\boldsymbol \xi}}
\def\kappav{{\boldsymbol \kappa}}  
\def\ii{\mbox{\rm i}}

\title{Contribution to the theory of waves in multi-dimensional linear dispersive media}

\author[V. G. Gnevyshev, and S. I. Badulin ]%
{V\ls L\ls A\ls D\ls I\ls M\ls I\ls R\ns G.\ns G\ls N\ls E\ls V\ls Y\ls S\ls H\ls E\ls V$^1$,%
\ns S\ls E\ls R\ls G\ls E\ls I\ns I.\ns B\ls A\ls D\ls U\ls L\ls I\ls N$^{1,2}$\thanks{Email address for correspondence: badulin.si@ocean.ru}}%

\affiliation{$^1$P.P. Shirshov Institute of Oceanology of the Russian Academy of Sciences, 36 Nakhimovsky pr., 117997, Moscow, Russia \\[\affilskip]
$^2$Skolkovo Institute of Science and Technology,  Bolshoy Boulevard 30, bld. 1, Moscow 121205, Russia }

\pubyear{2022}
\volume{XXX}
\pagerange{XXX--YYY}
\date{?; revised ?; accepted ?. - To be entered by editorial office}
\begin{document}

\maketitle

\begin{abstract}
The asymptotic solutions for linear waves generated by oscillating source of elliptic shape in the motionless media is constructed with the recently developed Reference Solution Approach (RSA). Pronounced anisotropy of the solutions is found for elongated sources both for amplitudes and phases of the resulting wave pattern. The classic Kelvin angles of the ship wave patterns determine specific directions of this anisotropy, thus, demonstrating the role of wave dispersion. {The analytical results within the RSA are shown to agree remarkably well with exact solutions of the linear wave problem.}
\end{abstract}


\section{Introduction}
In this paper we address the classic and, at the first glance, well-understood problem of linear dispersive waves in a multi-dimensional homogeneous media. The propagation of deep water gravity waves in two dimensions represents a good case for analysis mathematical and physical aspects of the general problem.

The solution for dispersive waves propagating in a homogeneous media can be presented as a superposition of partial  harmonics. The Fourier decomposition is generally used when these harmonics are simple sine or cosine functions
\begin{equation}\label{eq2.1}
 f_{n} ( \xv , t )=  \Real  (2\pi)^{-n/2}  \int F_{n}(\kappav)\cdot \exp[\mathbf{\rm{i}(\kappav \xv}- \omega(\kappav) t) ]
d\kappav
\end{equation}
Here the scalar function $f_n$ represents a scalar wave field variable, say, surface elevation of water surface waves or pressure perturbation of acoustic waves. In (\ref{eq2.1}) wavevector $\kappav$ and coordinate $\xv$ are vectors in $n-$dimensional infinite space, i.e. the integral limits in (\ref{eq2.1}) are assumed to be $\pm \infty$. Dispersion dependence $\omega(\kappav)$ specifies the particular type of wave motion. The associated effects of wave dispersion on evolution of the initial wave perturbation (\ref{eq2.1}) can be treated as interference of partial harmonics with the Fourier image amplitudes  $F_n$ and the corresponding phases of the exponent.

The integral (\ref{eq2.1}) cannot be explicitly estimated in the most of cases of interest. This is why asymptotic approaches such as the Stationary Phase Approximation (SPA) and the Steepest Descent Method (SDM) are widely used for the analysis of both quantitative and qualitative peculiarities of the wave evolution. The feature of the integrand in (\ref{eq2.1}) is used to get approximate expressions: the formally rapidly oscillating exponent makes contribution of just a very few domains of integration to be important. The resulting asymptotic expressions are inherently point-wise: they are determined by  special points of these domains and do not capture the non-local features of the wave field. In one-dimensional case, the SPA solution is determined by the group velocity, the first derivative of the dispersion relation or the gradient in the multi-dimensional space $\nabla_\kappav \omega(\kappav)$  and by the wave dispersion, the second derivatives $d^2\omega(\kappav)/d k_{i} d k_j$, $i,j=1\ldots n$) in the points of stationarity of the wave phase $\nabla_\kappav \omega(\kappav)=0$. The appearance of the second derivatives imposes certain difficulties for the use of SPA in the case of multi-dimensional wave propagation when the wave dispersion is treated simplistically. Multiplication of second derivatives, say, $\omega_{kk}\omega_{ll}$ in two-dimensional case ($\kappav=(k,l)$) looks logical and provides a correct law of dispersive decay $t^{-1}$. Vanishing one of the second derivatives uncover the deficiency of such treatment immediately. The singularity of the resulting SPA expressions, formally speaking, requires special consideration of terms of higher-order dispersion (derivatives of $\omega(\kappav)$ of the third and higher orders). In fact, the wave dispersion in multi-dimensional media is associated with the quadratic form of the second differential of the dispersion relation, the Hessian determinant \citep{Fedoryuk1994}. This characteristic is rotationally invariant in contrast to the above `intuitively derived' $\omega_{kk}\omega_{ll}$ as dispersion characteristics. \citet[sect.4.8][]{LighthillBook78} called this characteristic the Jacobian implying its remarkable association with transformation of wave volumes.

Kelvin himself faced the above difficulty when derived the corresponding angle $\cos^2\theta=2/3$ of the `singular' manifold in the problem of ship waves. The idea of Lord Kelvin `to blur, or to smooth it down' \citep[][p.425]{Kelvin1887OnShip} by accounting for finite size of wave source is physically transparent and has been realized in many studies, mostly, in the one-dimensional setups.

The inherently multi-dimensional approach of this paper, the Reference Solution Approach (RSA), in a sense, recapitulate the previous concepts. Another idea of  Lord Kelvin \citep[][\S~65, eq.95]{Kelvin1906} to use special family of initial conditions allows one to get analytical solutions  in explicit form  and give them a transparent physical interpretation.  \citet{Gnev2017} proposed to use the gaussian distribution as one possessing a remarkable property: the Fourier image of the exponent of quadratic form (gaussian function) is a gaussian function again \citep[see][]{Fedoryuk1987}. The property is valid for the case of multiple dimensions and does not provide singularities. Below we consider the two-dimensional case of deep water gravity waves as an example following results of recent papers \citep{Gnev2017,GnevBsi2020}. The  dispersion relation is well-known
\begin{equation}
\label{eq2.2}
 \omega(k,l) = \sqrt{g} (k^2+l^2)^{1/4}
\end{equation}
 where $k,l$ are $x-$ and $y-$components of the wavevector.

Starting with the gaussian initial condition
\begin{equation}\label{eq2.3}
  f_2(x,y,t=0)=      \frac{1}{ \Delta x  \Delta y}    \exp\left[ \ii k_0 x+\ii l_0 y -\frac{x^2}{2(\Delta x)^2} - \frac{y^2}{2(\Delta y)^2} \right]
\end{equation}
and its Fourier image
\begin{equation}\label{eq2.4}
  F_2(k,l)=\exp\left[ -\frac{(k-k_0)^2}{2(\Delta k)^2} - \frac{(l-l_0)^2}{2(\Delta l)^2} \right],
\end{equation}
with $\Delta k= (\Delta x)^{-1}$, $\Delta l= (\Delta y)^{-1}$ being the wave packet bandwidth we get a  solution at arbitrary time $t$ in accordance with general formula (\ref{eq2.1}). For formally small bandwidth $(\Delta_k^2+ \Delta_l^2)/(k^2+l^2) \ll 1 $ the integrand in (\ref{eq2.1}) can be approximated by gaussian function centered at the carrier wavevector $\kappav_0=(k_0,l_0)$ and the corresponding wave frequency $\omega_0=\omega(k_0,l_0)$. The integration in wavenumber space gives the gaussian-shaped wave packet in the coordinate space. Thus, the method does not lead to singularity.

In this paper we focus on the effect of the shape of finite size wave sources. This effect cannot be captured by the conventional asymptotic methods (e.g. SPA) which are inherently associated with point-wise sources. We develop the RSA analysis in an explicit analytic form and demonstrate somewhat paradoxical result. The wave perturbations near an elliptic source can be essentially anisotropic at distances essentially exceeding the source size. Moreover, these perturbations emphasize specific directions relatively to the main axes of the ellipse. These directions are associated with directions of minimal wave dispersion and the corresponding angles: Kelvin's angles.

In \S~2 we, first, recall key steps of the analysis within the RSA and present solutions for an oscillating motionless source of elliptic shape. The physical analysis is focused on pronounced features of anisotropy of the wave pulse at moderate distances from the source where conventional asymptotic methods like SPA are not formally valid. In this way, we demonstrate better performance of the RSA as compared to its predecessors and new, somewhat paradoxical, effect of wave dispersion.

Conclusions and Discussion in \S~3 is focused on details of the effect of wave dispersion in multi-dimensional media. Prospects of development of the paper approach for numerous problems of wave dynamics in different scopes of physics are also discussed.

\section{Linear wave propagation within the Reference Solution Approach}
The  asymptotic analysis within the Reference Solution Approach (RSA) can be split into two steps. First, calculate the center of the gaussian pulse, that is location of its maximum in the coordinate space. This center is associated with the carrier wave harmonic $\kappav_0$ in (\ref{eq2.4})  Secondly, the magnitude and the width of this pulse can be retrieved from the integration of (\ref{eq2.1}).

\subsection{The pulse center propagation: the point-wise wave packet kinematics}
The first step follows the well-known receipts:  the center of the pulse  $\xv_0=(x_0,y_0)$ propagates at the group velocity quite similarly to the SPA or geometric optics schemes
\begin{equation}
\label{eq2.5}
 x_{0}= \frac{\partial \omega}{\partial k} t, \qquad y_{0} = \frac{\partial \omega}{\partial l} t
\end{equation}
Using polar coordinates for wavevector $\kappav$
\begin{equation}\label{eq2.6}
  k_0=|\kappav_0| \cos \theta;\quad l_0=|\kappav_0| \sin \theta
\end{equation}
one can re-write (\ref{eq2.5}) as follows
\begin{equation}
\label{eq2.7}
 x_{0} =t \frac{\omega_0}{2\kappa_0} \cos \theta , \qquad  y_{0} = t \frac{\omega_0}{2\kappa_0} \sin \theta
\end{equation}
The Legendre transform $ \varphi = \omega  t - k x - l y $ specifies the phase in the center of the wave packet (the Lagrange function)
\begin{equation}\label{eq2.8}
  \varphi_{0}= \frac{1}{2} \omega_{0} t
\end{equation}
It leads to the parametric representation of the wave trajectories
\begin{equation}
\label{eq2.9}
 x_{0} = \frac{g \varphi_{0} }{\omega_{0}^{2} } \cos\theta; \quad
y_{0} =  \frac{g \varphi_{0} }{\omega_{0}^{2} }  \sin\theta
\end{equation}
and proposes natural dimensionless coordinates $\tilde x,\, \tilde y$
\begin{equation}
\label{eq2.10}
\tilde x_0 = x_0\frac{\omega_{0}^{2} }{g } ; \quad
\tilde y_0 =  y_0\frac {\omega_{0}^{2} }{g }
\end{equation}
The surface of the constant phase $\varphi_0={\rm const}$ is a circle
\begin{equation}\label{eq2.11}
  \varphi_0^2=\tilde x_0^2+\tilde y_0^2
\end{equation}
which radius is determined by the group velocity of the carrier harmonic (see eq.~\ref{eq2.7})
Below we use notations $x,\,y $ of dimensionless coordinates without tildes for $\tilde x,\,\tilde y$.
 We also leave similar notations for dimensionless wavevector and its components. One should note that in this part of our analysis we follow the conventional point-wise description of wave field of SPA or geometric optics.

\subsection{Wave packet dynamics. Wave dispersion effects}
Looking for the effects of finite size and shape of the waves source we consider initial conditions in the form (\ref{eq2.3},\ref{eq2.4}).
Following \citet{Gnev2017} one can present results of integration of (\ref{eq2.1}) with initial pulse (\ref{eq2.4}) as multiplication of three terms
\begin{equation}
\label{eq2.12}
F(\xiv,t)=  \left[D^{2}(t)\right]^{-1/4} \exp \left[ - \frac{C(\xiv,t)}{D^2(t)}\right]
 \cos \left[\varphi_0 (\xv_0,t) + \varphi_{1} (t) - \omega_2(\xiv,t)t \right]
 \end{equation}
Different arguments are introduced in (\ref{eq2.12}) to emphasize the solution structure. The `true' coordinate is split into two terms $\xv=\xv_0+\xiv$. The first term $\xv_0$ is a solution of the ray equations (\ref{eq2.5}) and, thus, describing the evolution of the pulse peak: the carrier wave frequency and wavenumber $\omega_0,\,\kappav_0$. The formally small coordinate $\xiv$ captures effects of small deviations from the carrier wave parameters.  The term
\begin{equation}
 \label{eq2.13}
 D^{2}(t) = \left[ 1 + t^{2}\left(\mu^{2}_{x y} - \mu_{x} \mu_{y}   \right)  \right]^{2} +
  t^{2} \left( \mu_{x} + \mu_{y}\right)^{2}
\end{equation}
can be naturally treated as the wave packet volume. Its presence in the exponent argument and in the pre-exponent guarantees conservation of the total energy of the pulse. {One should note that the physically essential feature of conservation is not so evident in the point-wise treatment of the problem (e.g. SPA).}

Transformation of the volume $D$ is associated with terms of wave dispersion
\begin{equation}
\label{eq2.14}
  \mu_{x} = ( \Delta k )^{2}\, \omega''_{k k }|_{\kappav=\kappav_0}; \quad
    \mu_{y} = ( \Delta l )^{2}\, \omega''_{l l }|_{\kappav=\kappav_0};  \quad
    \mu_{x y} =  ( \Delta k  \Delta l ) \, \omega''_{k l }|_{\kappav=\kappav_0},
\end{equation}
the second differentials of the series of dispersion relation (\ref{eq2.2}) in small deviation of the wave frequency from the carrier one $\omega(\kappav_0)$. The unity in the square brackets of (\ref{eq2.14}) secures validity of the approach at short times oppositely to the long-term asymptotics of SPA. It is easy to show that $D$ never vanishes and, thus, the novel approach is free of the problem of singularity. At infinitely long time, (\ref{eq2.14}) predicts conventional  SPA asymptotics $D \sim t^{-1}$ provided by the term in square brackets. However, a new term in the left hand side appears as compared to SPA. The vanishing of the combination of the second differentials $   \mu_{x}\mu_{y} -\mu_{xy}^2= 0 $ makes this term the key one at long times and the wave pulse volume $D$ decays slower as $ t^{-1/2}$. This effect has been presented by \citet{Gnev2017} as \emph{quasi-dispersion}. In general case, the new term is important at relatively small time which is not covered by the SPA.

The novel approach of RSA adds more features of propagation by introducing  the pulse shape explicitly. This shape remains gaussian with a quadratic form  in the argument of the exponent (\ref{eq2.12})
\begin{equation}\label{eq2.15}
   C(\xi,\eta)=t^2\left[(\Delta k \xi \mu_{xy} - \Delta l \eta \mu_x)^2 + (\Delta l \eta \mu_{xy}- \Delta k \xi \mu_y)^2\right] + (\Delta k)^2 \xi^2 + (\Delta l)^2 \eta^2,
 \end{equation}
The formally small deviations $\xi=x-x_0,\, \eta=y-y_0$ from the pulse center $\xv_0$ in the coordinate space (similarly in wavevector space) afford new manifestations of the effect of wave dispersion. The elliptic pulse can, first, rotate and, secondly, change its eccentricity. At large times (distance from the source), main directions of the ellipse $C(\xi,\eta)={\rm const}$ (\ref{eq2.15}) correspond to directions of the fastest and the slowest wave dispersion accounting for the wave packet bandwidth (i.e. productions of the second derivatives of dispersion relation (\ref{eq2.2}) on bandwidth parameters $\Delta k,\,\Delta l $ in the spirit of (\ref{eq2.14}).

The first phase function in (\ref{eq2.12})
\begin{equation}
\label{eq2.16}
  \varphi_0=\kappav_0\xv - \omega_0 t
\end{equation}
is just a phase of the carrier wave with wavevector $\kappav_0$ and frequency $\omega_0$ quite similarly to the SPA approach. The second phase function
\begin{equation}
\label{eq2.17}
\varphi_{1} (t)=  - \frac{1}{2}\, \arctan \left[
 \frac{ t (\mu_{x} + \mu_{y})      }
 { 1 + t^{2} ( \mu^{2}_{xy} - \mu_{x}\mu_{y})  }
 \right]
 \end{equation}
 can also be naturally related to the result of the SPA. In the uni-directional case the dependence on time vanishes in the denominator of argument in (\ref{eq2.17}) and the phase $\varphi_1$ is tending asymptotically to the well-known phase shift $\pm \upi/4$. Our approach (RSA) predicts non-trivial dependence on time in general case and, additionally, discovers the origin of the phase shift: wave dispersion. The last phase function $\omega_2(\xiv,t) t$ is also associated with the dispersion of wave harmonics relatively to the carrier one $\kappav_0$. It introduces a correction to the carrier frequency
  \begin{eqnarray}
  \omega_2(\xiv,t)& = & 2\left[\Delta k^2 (t^2\mu_y(\mu_{xy}-\mu_x\mu_y)-t\mu_x) \xi^2 + \Delta l^2(t^2\mu_x(\mu_{xy}-\mu_x\mu_y)-t\mu_y)\eta^2  \right. \nonumber\\ \label{eq2.18}
 & + & \left. 2\Delta k \Delta l  t \mu_{xy}(1+t^2(\mu_{xy}-\mu_x\mu_y)) \xi \eta \right] \cdot D^{-2},
 \end{eqnarray}
 which is  a time-dependent quadratic form of small deviations from the wave pulse center $\xv_0$ quite similarly to the argument of the gaussian pulse shape $C(\xiv,t)$ (\ref{eq2.15}). This term is, evidently, plain zero in the pulse center $\xv=\xv_0$ that is for the carrier harmonic and grows to the periphery of the wave pulse, thus, reflecting the effect of wave dispersion.  Again, the term $\omega_2$ describes rotation and stretching-squeezing of the isolines of constant phase corrections. At large time $\omega_2 $ is vanishing as $t^{-2}$.

 \subsection{Waves from an elliptic source within RSA}
Strictly speaking, the RSA is valid under assumption of small wave packet bandwidth $\left[(\Delta k)^2+ (\Delta l)^2\right]/(k^2+l^2) \ll 1 $. In fact, it does not impose essential restrictions for the method extension and construction of more general solutions as soon as the principle of linear superposition continues to work. More sophisticated solutions can be constructed as combinations of the narrow-banded elements.

Let us consider a family of initial conditions (\ref{eq2.3},\ref{eq2.4}) with carrier wavevectors $\kappav_0$ with fixed modulo and directions in the range $-\upi \le \theta \le \upi$.  In this way, solution (\ref{eq2.12}) parameterized with the direction (wavevector angle $\theta$) becomes a solution for the source radiating in the whole range of angles.  The restriction on bandwidth becomes unessential in absence of interference of the partial solutions with narrow bandwidth.

 The RSA proposes quite sophisticated structure of wave solutions (\ref{eq2.12}). Here we consider `essential' features by neglecting dependencies on local (relatively to the pulse center) coordinate $\xiv$. First, the  dependence $D$ on time (\ref{eq2.13}) or the pulse coordinate $\xiv_0$ describes the pulse envelope. Simple algebra gives for dispersion parameters of deep water waves
 \begin{eqnarray}
\label{eq2.19}
 \mu^{2}_{x y} - \mu_{x} \mu_{y}    & = &
  \frac{(\Delta k )^{2} (\Delta l )^{2}}{8} \frac{g^{4}}{\omega_{0}^{6}} \\
 \label{eq2.20}
 \mu_{x} & = & \frac{(\Delta k )^{2}}{2} \frac{g^2}{\omega_0^3}(1- \frac{3}{2}\cos^{2}\theta ) \\
\label{eq2.21}
 \mu_{y} & = & \frac{(\Delta l )^{2}  }{2 }\,  \frac{g^{2}}{\omega_{0}^{3}} (1- \frac{3}{2} \sin^{2} \theta ).
\end{eqnarray}
Two conjugate angles $\cos^2\theta_{K1}=2/3$ and  $\sin^2\theta_{K2}=2/3$ when expressions (\ref{eq2.20},\ref{eq2.21}) vanish are the well-known angles of Kelvin \citep{Kelvin1887OnShip} relatively to the main axes of the elliptical source. Plugging (\ref{eq2.19}-\ref{eq2.21}) into (\ref{eq2.13}) with  $\varphi_0$ as an argument of wave evolution gives (see \ref{eq2.8})
\begin{equation}
\label{eq2.22}
D^{2} = \left(1+A^2/2\right)^2 +B^2
 \end{equation}
 with
 \begin{eqnarray}
 \label{eq2.23}
   A &=&\varphi_0  \frac{\Delta k \Delta l}{\kappa^2};
     \\ \label{eq2.24}
   B &= & \varphi_0 \left[  (1- \frac{3}{2} \cos^2 \theta )\left(\frac{\Delta k}{\kappa} \right)^2 + (1- \frac{3}{2} \sin^2 \theta ) \left(\frac{\Delta l}{\kappa} \right)^2 \right].
  \end{eqnarray}
 One should note that all these expressions do not depend on wave parameters but only on the bandwidth $\Delta k/\kappa,\,\Delta l/\kappa$. The term $A$ in (\ref{eq2.22}) gives an asymptotic dependence at large time $t\to \infty $ as $D \sim \varphi_0^2$ that corresponds to the  magnitude decay $t^{-1}$. This decay comes from two physical effects: the wave dispersion as $t^{-1/2}$ along the direction of propagation and the cylindrical divergence with the same law of decay. The latter can be naturally treated  as the same wave dispersion in transversal direction. In this way, the RSA is completely consistent with the conventional SPA or geometric optics approaches. The shape of the source is unessential in this case. The multiplier  $\Delta k \times \Delta l$ in (\ref{eq2.23}) is just an area of the pulse in the wavenumber space or, alternatively, the inverse area of the pulse in the coordinate space.

  The second term in (\ref{eq2.22}) $B$ represents key differences of the RSA and the conventional approaches. First, it is proportional to the phase function $\varphi_0$ and, thus, becomes vanishingly small at large time as compared to $A$. The term $B$ can contribute significantly at relatively short time (distance) when SPA is not valid but the novel RSA adequately works.
  Secondly, the term $B$ is inherently anisotropic. Depending on the orientation of the elliptical source this term can emphasize one of the Kelvin angles $\theta_{K1}$ or $\theta_{K2}$.
 \begin{figure}
  \centering
  a) \hskip 5 cm b)\\
  \includegraphics[scale=0.45]{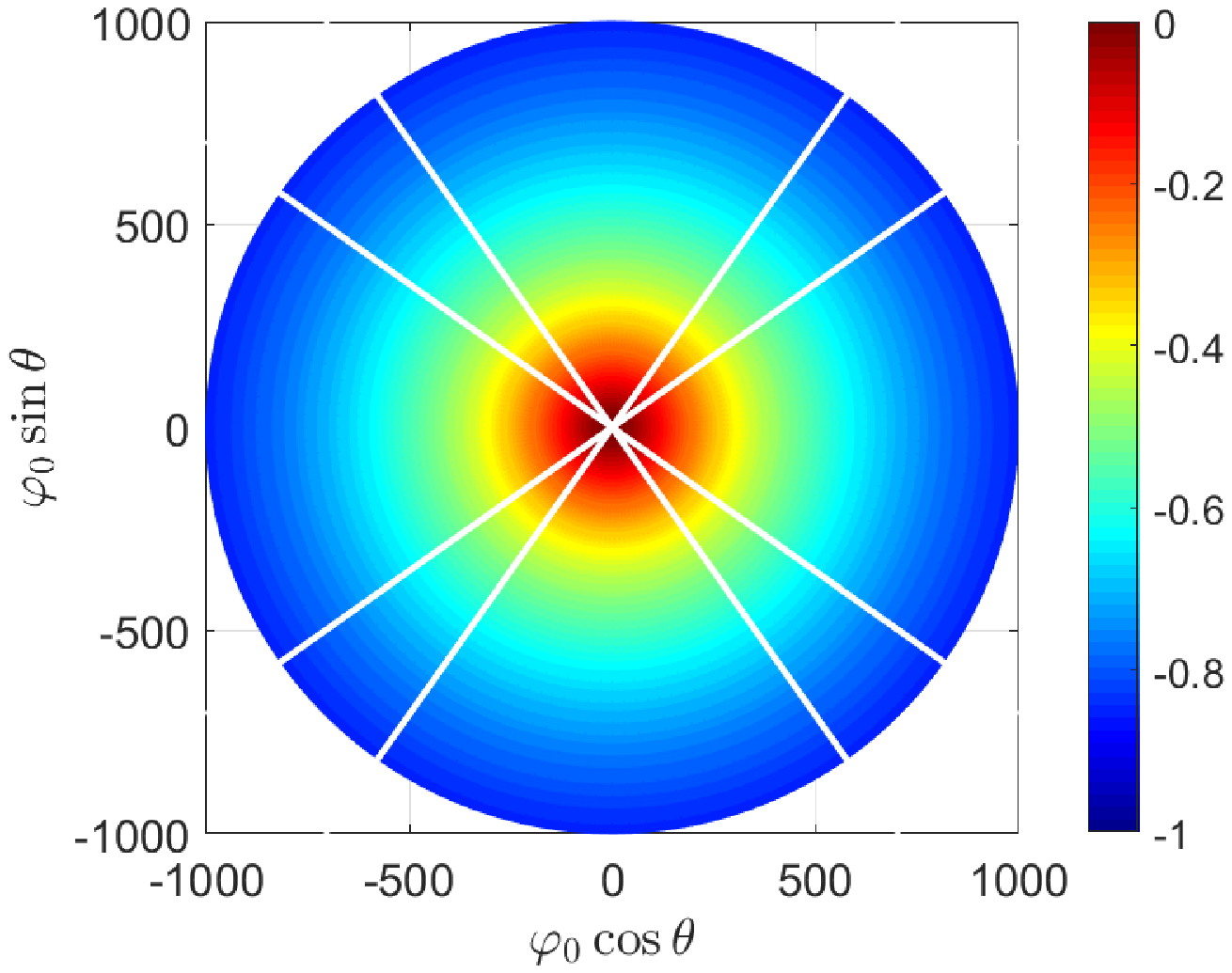}
  \includegraphics[scale=0.45]{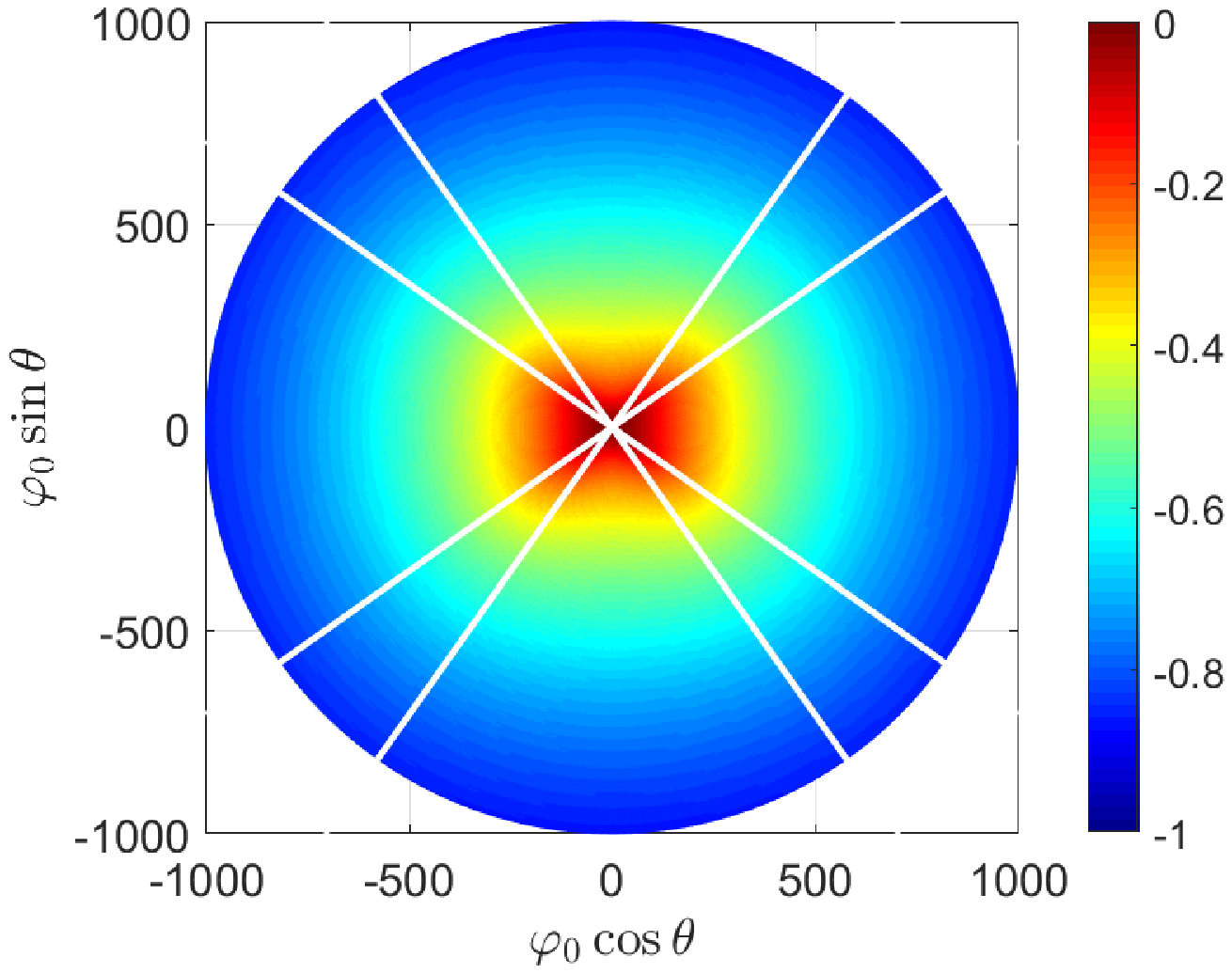}
  c) \hskip 5 cm d)\\
  \includegraphics[scale=0.45]{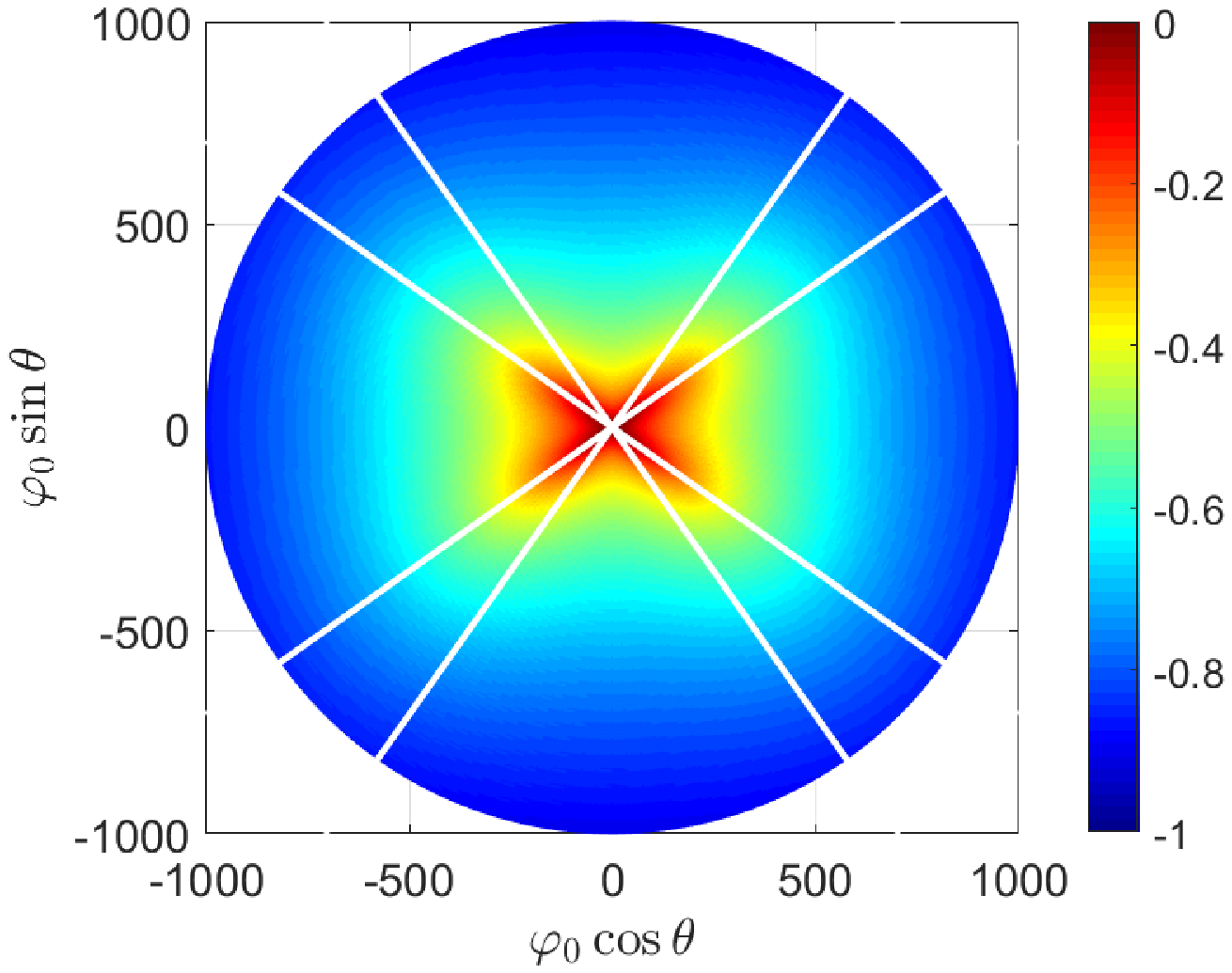}
  \includegraphics[scale=0.45]{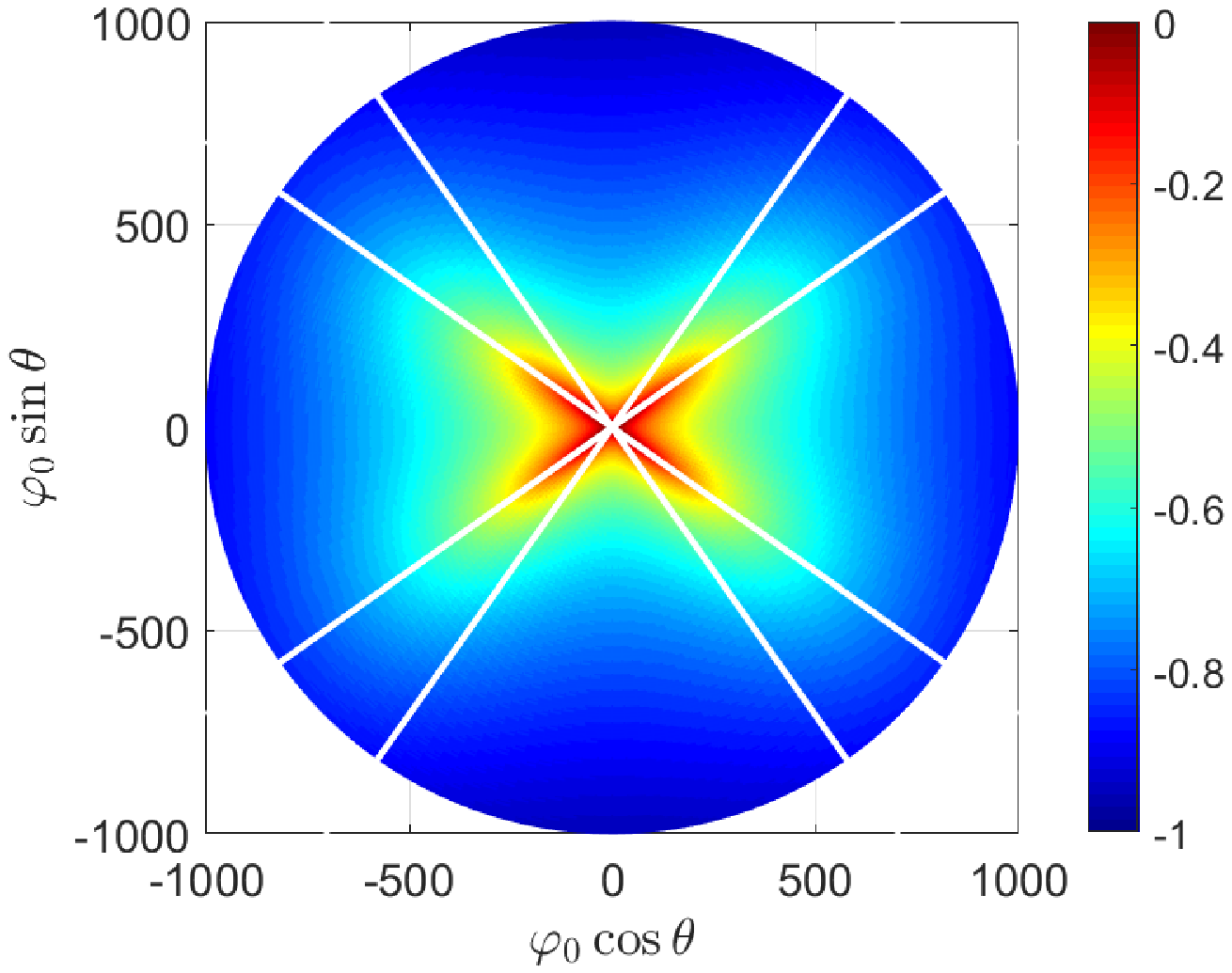}
        \caption{ The amplitude (decimal logarithm) of envelope of cylindrical wave pulse from elliptic sources with different aspect ratios in accordance with the RSA result (\ref{eq2.12},\ref{eq2.22}). White lines show Kelvin's angles. The pulse area $\Delta k \Delta l/\kappa^2=0.01$ is fixed.   a) -- circular source $\Delta k/\Delta l = 1$; b) -- $\Delta k/\Delta l = 2.25 $; c) -- $\Delta k/\Delta l =4$; d) -- $\Delta k/\Delta l =6.25$}
  \label{fig01}
\end{figure}

   Figure~\ref{fig01} illustrates the effect of anisotropy of wave perturbations from the source of elliptical shape. We fix the ratio $\Delta k \Delta l/\kappa^2= 0.01 $, the pulse square, and then vary the ratio $\varepsilon=\Delta k/\Delta l$, the aspect ratio of the initial pulse (\ref{eq2.4}).  The circular source, evidently, produces a symmetric pattern (figure~\ref{fig01}{\it a}). The growing eccentricity breaks this symmetry. Elongated sources (we show $\varepsilon = 2.25,\, 4, \, 6.25$) form an oblique cross with the double Kelvin opening angle $2\theta_{K1}= 2 \arccos(2/3) \approx 70.53^\circ$ which is supplementary angle to the classic Kelvin angle of the stationary ship wave wake on deep water $\theta_K=19.47^\circ$. Figure~\ref{fig01} shows the decimal logarithm of the wave envelope magnitude for the radii $\varphi_0 < 1000$ that is for number of wave periods $2000/\upi$ or $1000/\upi$ carrier wavelengths. At these scales the pulse becomes almost isotropic that is the isotropic `SPA term' $A$ becomes dominant as compared to the `non-SPA' $B$ in (\ref{eq2.22}).
 \begin{figure}
  \centering
  a) \hskip 5 cm b)\\
  \includegraphics[scale=0.45]{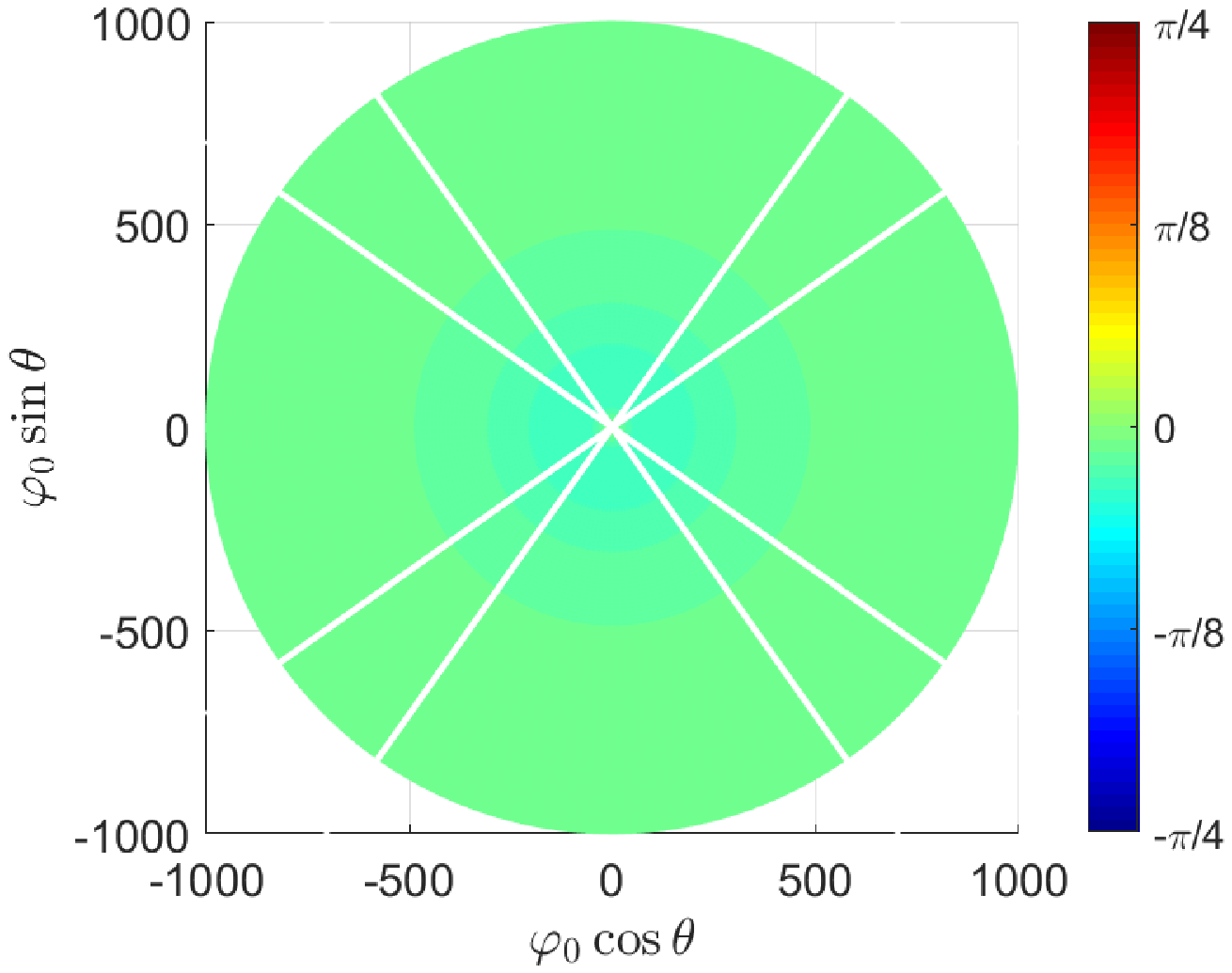}
  \includegraphics[scale=0.45]{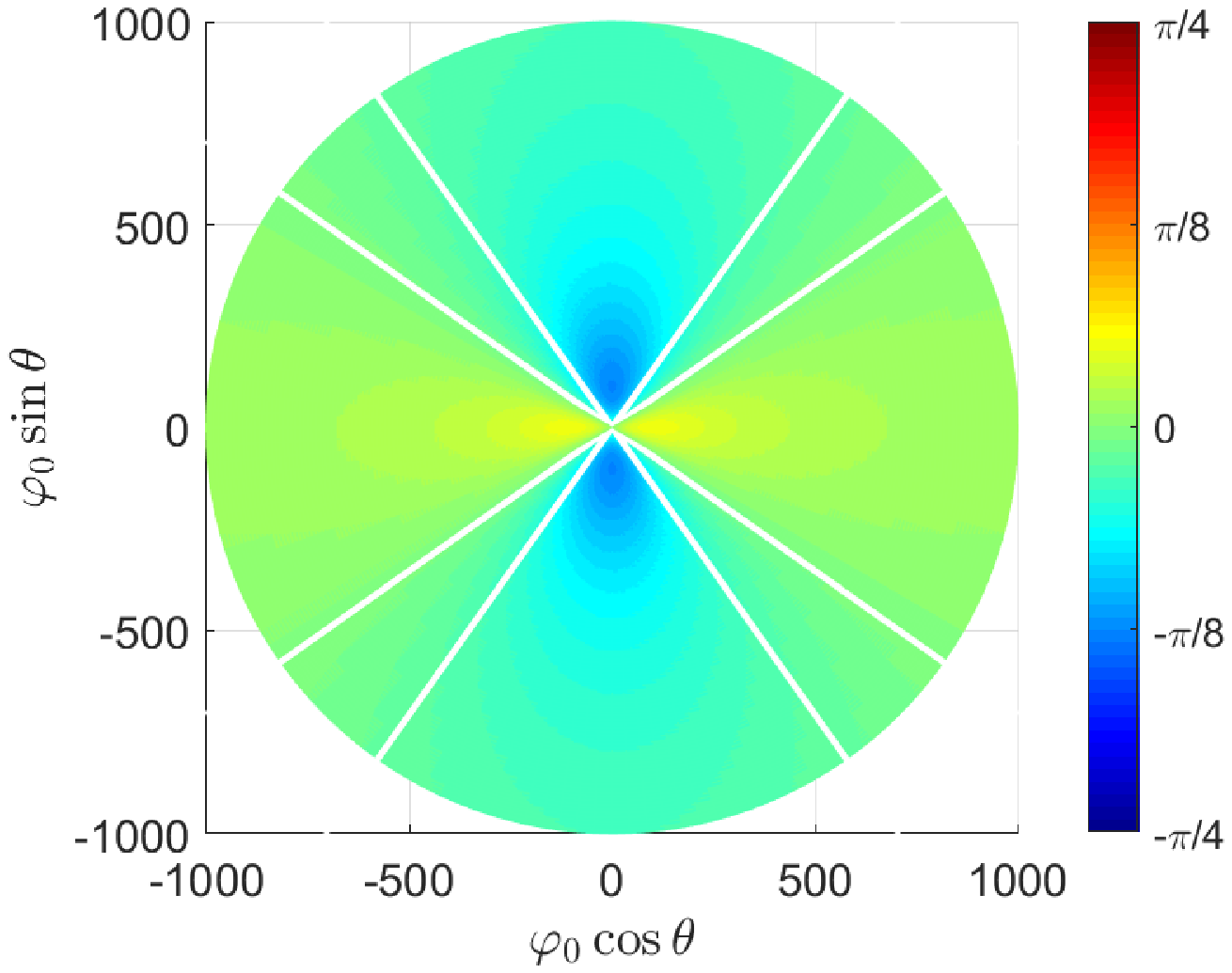}
  c) \hskip 5 cm d)\\
  \includegraphics[scale=0.45]{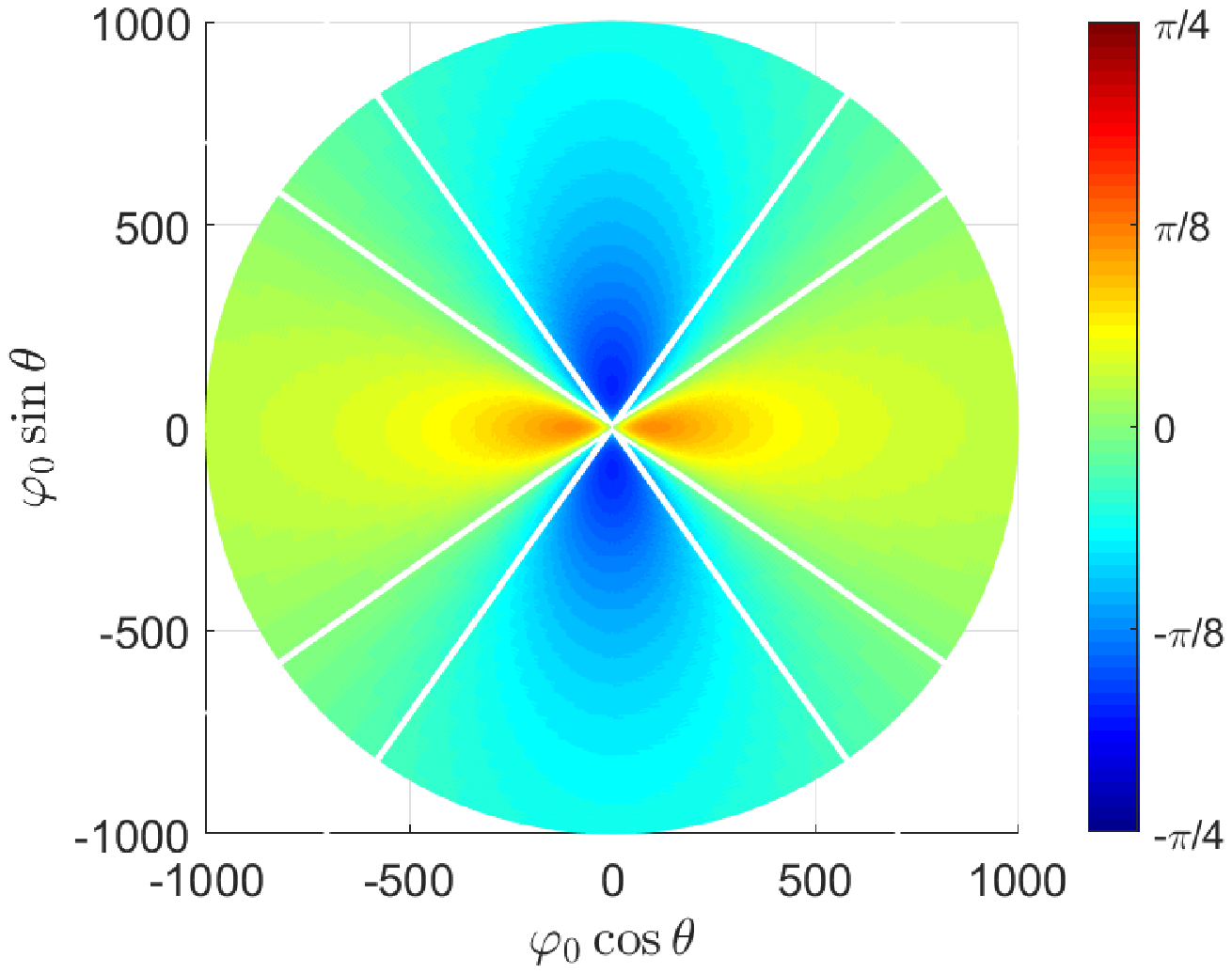}
  \includegraphics[scale=0.45]{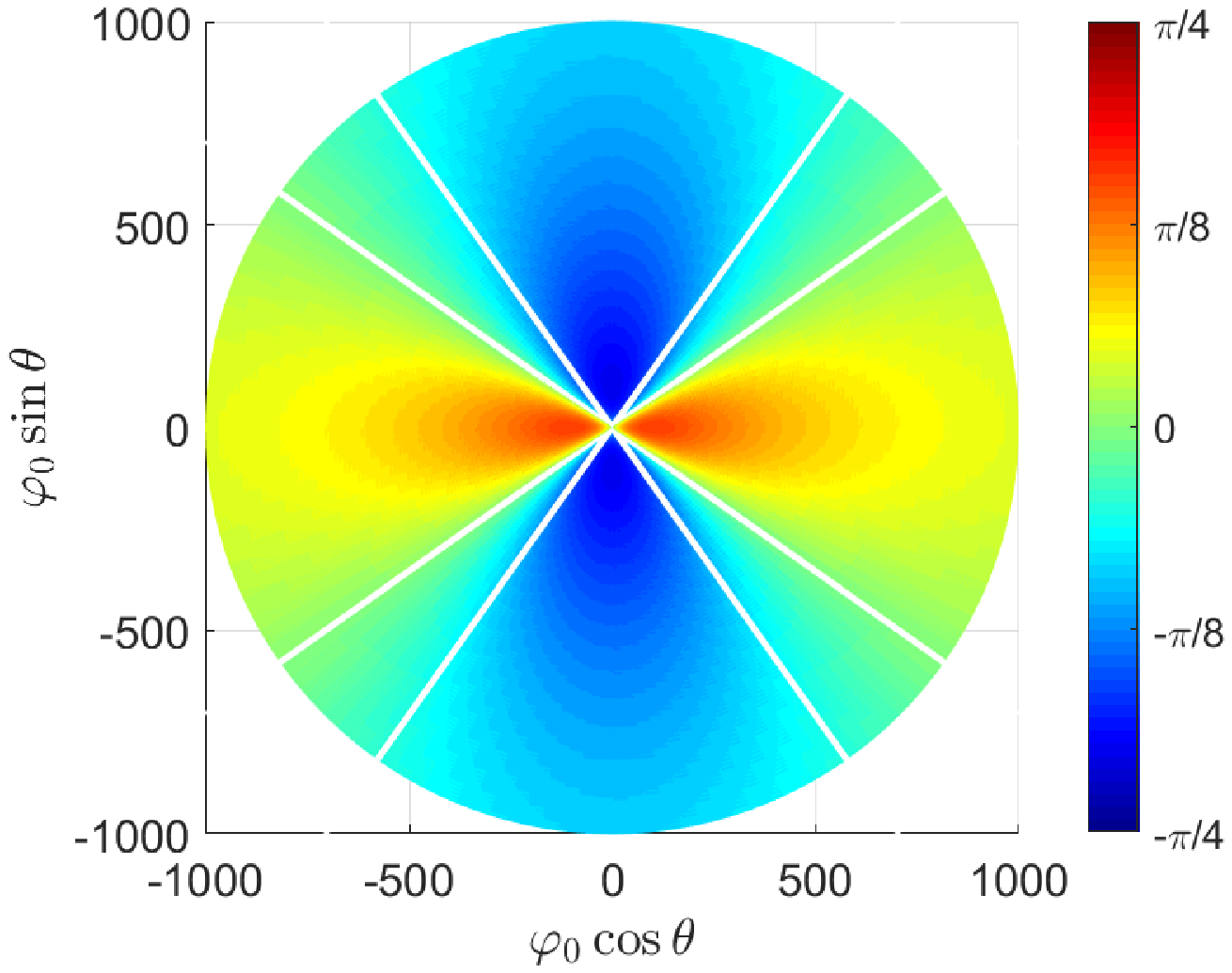}
        \caption{The phase function $\varphi_0$ (\ref{eq2.25}) of cylindrical wave pulse from elliptic sources with different aspect ratios in accordance with the RSA result (\ref{eq2.12},\ref{eq2.22}).  White lines show Kelvin's angles. a) -- circular source $\Delta k/\Delta l = 1$; b) -- $\Delta k/\Delta l = 2.25 $; c) -- $\Delta k/\Delta l =4$; d) -- $\Delta k/\Delta l =6.25$.}
  \label{fig02}
\end{figure}

The anisotropy of the pulse from elliptic source is also seen in the phase function $\varphi_1$ (\ref{eq2.17}) that can be re-written in terms of the above functions $A,\,B$
\begin{equation}
\label{eq2.25}
\varphi_{1} (x,y)=  - \frac{1}{2}\, \arctan \left[ \frac{ B }{ 1+   A^2 } \right].
 \end{equation}
 The results are shown in Figure~\ref{fig02} for the same set of parameter $\varepsilon=\Delta k/\Delta l$ as in Figure~\ref{fig01}. One can see pronounced variations of function $\varphi_1$ in the range of the reference values of the SPA $\pm \upi/4 $. These variations show slower tendency to isotropisation of solutions with distance as compared to those for amplitudes in Figure~\ref{fig01}.

\subsection{Waves from an elliptic source within direct simulation}
\begin{figure}
  \centering
  a) \hskip 4.5 cm b)\\
  \includegraphics[scale=0.45]{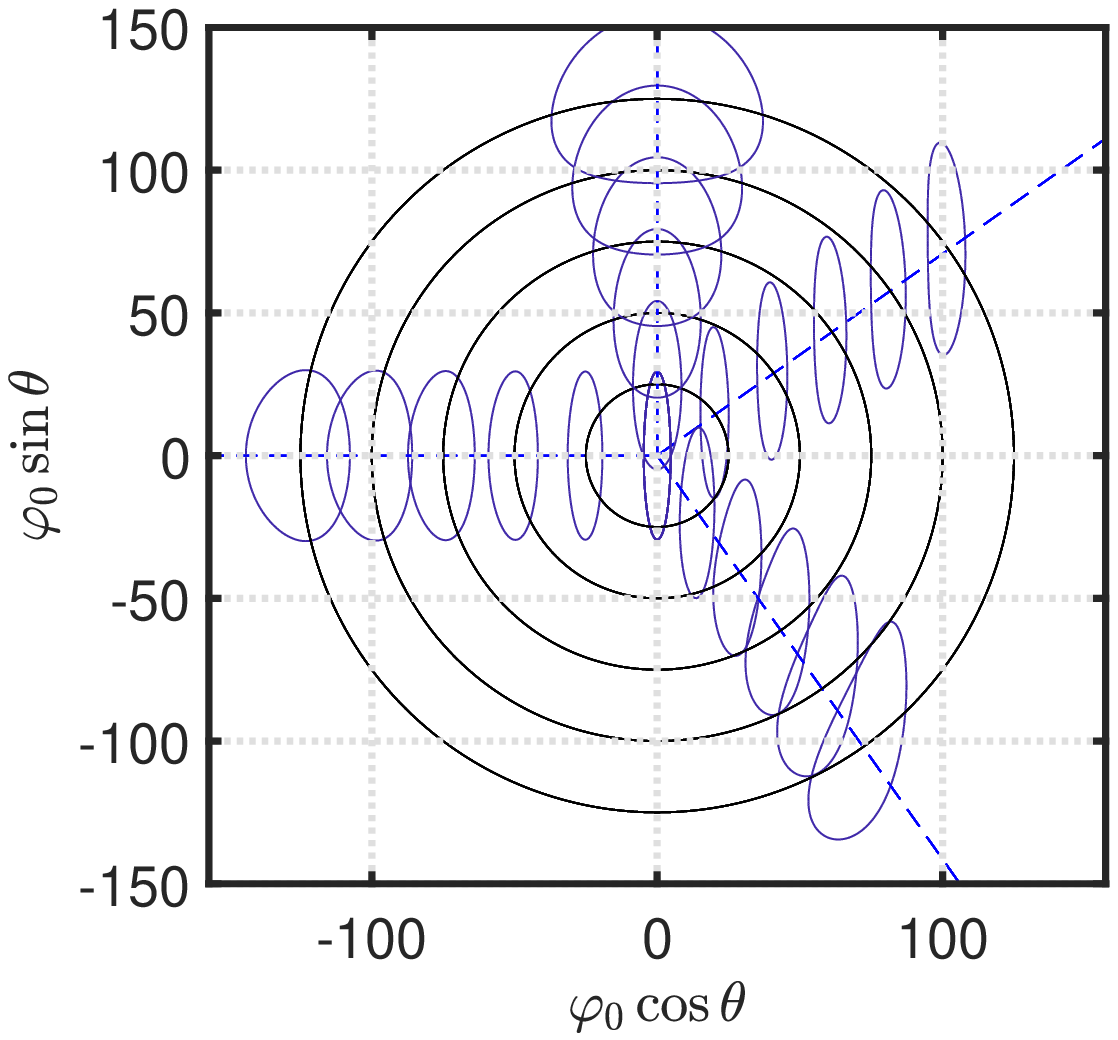}
  \includegraphics[scale=0.45]{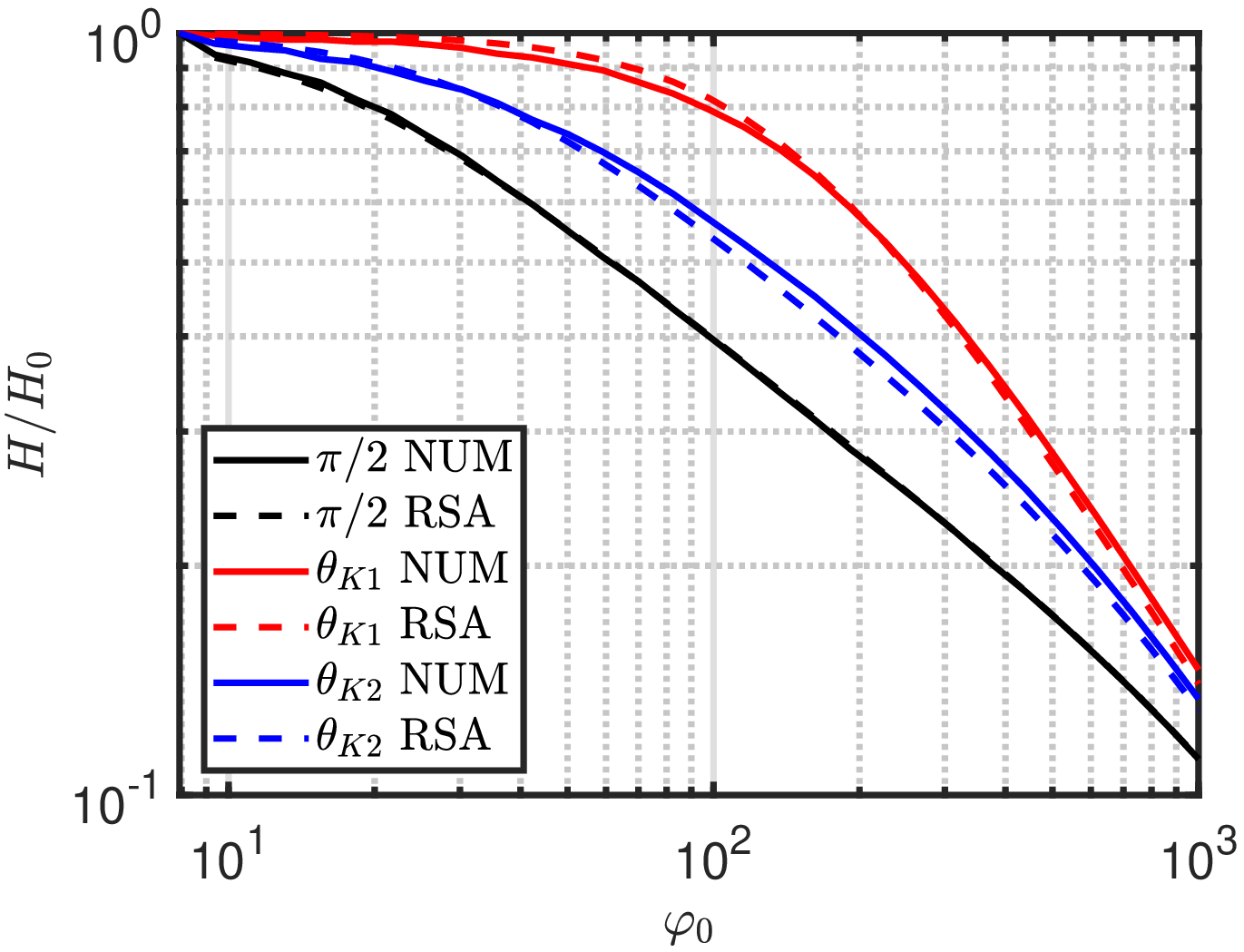}
          \caption{{\it a)} -- Evolution of narrow-banded pulses ($\varepsilon  = \Delta k/\Delta l =6.25$) for different directions relatively to the ellipse main axes. Dashed lines -- pulse trajectories for angles $\theta= \pi/2,\, \theta_{K1},\, \theta_{K2}$. Circles correspond to $\varphi_0=25,\,50,\,75,\,100,\,125$. {\it b)} -- The normalized envelope amplitude for different directions (see legend) as function of $\varphi_0$. Solid line -- direct simulation with (\ref{eq2.1}, dashed -- the RSA results.}
  \label{fig03}
\end{figure}

The validity of simple analytic expressions provided by RSA is an important issue of the study. For the problem discussed the problem can be easily resolved by direct integration of the basic relationship for the linear waves (\ref{eq2.1}). We do this for elementary narrow-banded solutions as they are treated by the RSA (\ref{eq2.3},\ref{eq2.4}). Figure~\ref{fig03}{\it a} shows evolution of initial elliptic perturbation in time (distance) for different directions relatively to the ellipse main axes. The case corresponds to Figures~1{\it d}, 2{\it d} of the maximal elongation $\varepsilon=\Delta k/\Delta l=6.25$. The figure~\ref{fig03}{\it a} shows  shapes as the pulse half-width. One can see rather strong dispersion of pulses along all the directions but $\theta_{K1}$ relatively to the $x-$axis. One can observe the pronounced effect of the pulse rotation for the angle $\theta_{K2}$. The shapes of the pulses at long times differ essentially from the ellipses. Nevertheless, the evolution of the pulse magnitudes as the ration of the current amplitudes to an initial value follows remarkably well to the RSA prediction (\ref{eq2.12}).

In Figure~\ref{fig03}{\it b} the evolution of the pulse magnitude $H/H_0$ ($H_0=1/(\Delta x \Delta y)=f_2(x=0,y=0,t=0)$ in Eq.~\ref{eq2.3}, $H$ is the instant magnitude of the evolving pulse in Eq.~\ref{eq2.1}) is shown in comparison with its RSA counterpart $D^{-1/2}$ in (\ref{eq2.12}). The case $\theta=0$ (along $x-$ axis) is not shown because it coincides almost perfectly with the case $\theta_{K2}$. The curves are qualitatively consistent with tracing wave packets in Figure~\ref{fig03}{\it a}: the weakest blurring occurs along the Kelvin angle $\theta_{K1}$. This figure show an impressive matching of the RSA results with direct calculations of the integral (\ref{eq2.1}). The validity of RSA at short times is not surprising: the approximation works well when the pulse is close to the gaussian shape. But the approach continues to be valid for longer times when the pulse shape deviates from the reference shape as it is clearly seen from Figure~\ref{fig03}{\it a}. An additional argument for the RSA is its consistency with SPA asymptotics at large times.

Summarizing this section one can conclude that the effect of the wave source shape is captured by the RSA quite well both quantitatively and analytically as it is illustrated by Figures~\ref{fig01}-\ref{fig03}.

\section{Conclusions and Discussion}
 The finite size of wave source is the key point of the Reference Solution Method we develop in this paper. The conventional asymptotic methods, generally, work in `the far zone' where waves `forget' details of initial conditions. The latter can be treated as the problem of a point-wise source. We develop the Reference Solution Approach that accounts for a finite size of the source and, more, operates with the shape of the source in explicit analytical form. The family of the linear wave solutions in the form of the gaussian-shaped pulses in the wavevector space does not imply restrictions on the distance from the source and, potentially, is able to capture the effects of wave propagation at small distances. The accuracy of the RSA at large distances can be questionable when wave dispersion strongly corrupts the initially gaussian shape of the pulse in the coordinate space (cf. Figure~\ref{fig03}{\it a}). It is unlikely our case where significant  deviations from elliptic shapes Figure~\ref{fig03}{\it a}  of wave packets do not lead to visible quantitative effects in Figure~\ref{fig03}{\it b}.

      This paper presents solid arguments for the validity of the RSA, first, through comparison of this approach with direct integration of (\ref{eq2.1}) as a general solution for the problem of linear waves (see Figure~\ref{fig03}{\it b}). We discover pronounced effects of the source shape at distances up to $200$ wave periods at least where the inherently point-wise method of SPA cannot be applied. At larger distances one can see isotropisation of solutions in terms of wave amplitudes. Moreover, similar effect can be observed for wave phases that also reflect the effect of the source shape (see Figure~\ref{fig02}). Within the RSA the phase shift is a continuous function (\ref{eq2.25}) explicitly associated with wave dispersion through the dispersion characteristics (\ref{eq2.19}-\ref{eq2.21}). The physical origin of this shift is absolutely clear in contrast to the SPA where the nature of this characteristic is hidden by formal mathematical criteria.

 The effect of the source shape is manifested by specific pattern of the Kelvin's cross (angles $\theta_{K1},\,\theta_{K2}$) and can be easily treated as one built into the dispersion relation (\ref{eq2.1}) or similar power-law dependencies for other types of waves of negative dispersion.

 The Kelvin angle is usually associated with ship waves \citep{Kelvin1887OnShip} but not with motionless sources as in our case. At one hand, our work breaks this stereotype. At the other hand, it discovers new problems where the shape of wave patterns provokes essential physical effects.

 The problem of ship waves is generally treated as one of a stationary wave wake where only a special subset of wave harmonics is taken into account. These harmonics have zero frequency in the reference system of the wave source
 \begin{equation}\label{eq3.1}
   \omega(\kappav)+\kappav \Uv=0,
 \end{equation}
 i.e. do not propagate, stay stationary relatively to the source in the flow with the current speed $\Uv$. A natural question arises: is the solution on the manifold (\ref{eq3.1}) structurally stable? In other words, can  small change  of the problem setup provoke great changes of the resulting wave pattern? This study presupposes a setup to get answer on the structural stability challenge. The stationary wave wake pattern can be inspected as a limit of weakly non-stationary wave pattern where the right-hand side of (\ref{eq3.1}) is not zero but a small parameter. Within the close setup \citet{Lighthill1965} demonstrated features of the transition to the limit of stationary deep water wave wake.  \citet{BulatovVladimirov2019b} extended this analysis by constructing uniform asymptotics for the problem with two parameters: the current speed $\Uv$ and a frequency $\omega$ of the oscillating point-wise source. Essential physical effects, new branches of the wake patterns, can be found for other types of waves like planetary \citep{Lighthill1965} or internal waves \citep[e.g.][]{BulatovVladimirov2019a}. One can conclude that the simplest model presented in this paper can be regarded as a starting point for new prospective studies of wave dynamics.

\begin{acknowledgments}
 The work is supported by Russian Science Foundation Grant $\sharp$19-72-30028 and by the State Assignment $\sharp$0128-2021-0003. The authors appreciate the continuing support of the agencies.
\end{acknowledgments}


\begin{thebibliography}{10}
\expandafter\ifx\csname natexlab\endcsname\relax\def\natexlab#1{#1}\fi
\def\au#1{#1} \def\ed#1{#1} \def\yr#1{#1}\def\at#1{#1}\def\jt#1{\textit{#1}}
  \def\bt#1{#1}\def\bvol#1{\textbf{#1}} \def\vol#1{#1} \def\pg#1{#1}
  \def\publ#1{#1}\def\arxiv#1{#1}\def\org#1{#1}\def\st#1{\textit{#1}}

\bibitem[Bulatov \& Vladimirov(2019)]{BulatovVladimirov2019a}
{\sc \au{Bulatov, V.~V.} \& \au{Vladimirov, Yu.~V.}} \yr{2019}  \at{Internal
  gravity waves from a moving source: modeling and asymptotics}.  \jt{J. Phys.:
  Conf. Ser.} ~(1268),  \pg{012013}.

\bibitem[Bulatov \& Vladimirov(2019b)]{BulatovVladimirov2019b}
{\sc \au{Bulatov, V.~V.} \& \au{Vladimirov, Yu.~V.}} \yr{2019b}  \at{Far
  surface gravity waves fields under unstable generation regimes}.  \jt{J.
  Phys.: Conf. Ser.} ~(1392),  \pg{012005}.

\bibitem[Fedoryuk(1987)]{Fedoryuk1987}
{\sc \au{Fedoryuk, M.~V.}} \yr{1987} {\em Asymptotic: Integrals and Series\/}.
  \publ{Moscow: Nauka}.

\bibitem[Fedoryuk(1994)]{Fedoryuk1994}
{\sc \au{Fedoryuk, M.~V.}} \yr{1994} {\em Encyclopedia of Mathematics\/}, chap.
  Saddle point method.  \publ{Springer Science+Business Media B.V. / Kluwer
  Academic Publishers}.

\bibitem[Gnevyshev \& Badulin(2020)]{GnevBsi2020}
{\sc \au{Gnevyshev, V.} \& \au{Badulin, S.}} \yr{2020}  \at{Wave
  patterns of gravity-capillary waves from moving localized sources}.
  \jt{Fluids}  \bvol{5}~(4).

\bibitem[Gnevyshev \& Badulin(2017)]{Gnev2017}
{\sc \au{Gnevyshev, V.~G.} \& \au{Badulin, S.~I.}} \yr{2017}  \at{On
  the asymptotics of multidimensional linear wave packets: Reference
  solutions}.  \jt{Moscow University Physics Bulletin}  \bvol{72}~(4),
  \pg{415--423}.

\bibitem[Kelvin(1906)]{Kelvin1906}
{\sc \au{Kelvin, Lord}} \yr{1906}  \at{Deep sea ship waves}.  \jt{Proceedings
  of the Royal Society of Edinburgh}  \bvol{25}~(2),  \pg{1060--1084}.

\bibitem[Lighthill(1978)]{LighthillBook78}
{\sc \au{Lighthill, J.}} \yr{1978} {\em Waves in fluids\/}.  \publ{Cambridge,
  United Kingdom: Cambridge University Press}, 504 p.

\bibitem[Lighthill(1965)]{Lighthill1965}
{\sc \au{Lighthill, M.~J.}} \yr{1965}  \at{Contributions to the theory of waves
  in nonlinear dispersive systems}.  \jt{J. Inst. Maths. Appl.} ~(1),
  \pg{269--306}.

\bibitem[Thomson(1887)]{Kelvin1887OnShip}
{\sc \au{Thomson, W.}} \yr{1887}  \at{On ship waves}.  \jt{Proc. Inst.
  Mech. Engrs}  \bvol{38},  \pg{409--434}.

\end{thebibliography}

\end{document}